%% file: main.tex
\documentclass{llncs}

\usepackage[utf8]{inputenc}
\usepackage[T1]{fontenc}
\usepackage{xspace}
\usepackage{stmaryrd}
\usepackage{./why3lang}
\usepackage{./gospel}
\usepackage{./lstcoq}
\usepackage{color, colortbl}
\usepackage{graphicx}
\usepackage{array}
\usepackage{cellspace}
\usepackage{cite}
\usepackage{ latexsym }
\usepackage{subcaption}
\usepackage[misc]{ifsym}

\input{./mymacros.tex}

\usepackage{hyperref}

\def\_{\kern.08em\vbox{\hrule width.35em height.6pt}\kern.08em}

\newcommand{\cameleer}{\textsf{Cameleer}\xspace}
\newcommand{\GOSPEL}{{\textsf{GOSPEL}}\xspace}
\newcommand{\pre}{precondition\xspace}
\newcommand{\post}{postcondition\xspace}

\newcommand{\nbnote}[3]{
  \fcolorbox{gray}{yellow}{\bfseries\sffamily\scriptsize#1}
  {\color{#2} \sffamily\small$\blacktriangleright$\textit{#3}$\blacktriangleleft$}
}

\newcommand{\mario}[1]{\nbnote{Mario}{red}{#1}}
\renewcommand{\mario}[1]{}

\definecolor{thegray}{rgb}{0.849,0.849,0.849}

\usepackage{tikz}
\usetikzlibrary{arrows, shapes.misc, positioning, shapes.geometric}
\tikzset{>=stealth'}

\begin{document}

\title{Practical Deductive Verification of\\OCaml Programs (Extended Version)\thanks{This
    work is partly supported by Agence Nationale de la Recherche (ANR)
    grant ANR-22-CE48-0013-01 (GOSPEL) and NOVA LINCS
    ref. UIDB/04516/2020
    (\url{https://doi.org/10.54499/UIDB/04516/2020}) and
    ref. UIDP/04516/2020
    (\url{https://doi.org/10.54499/UIDP/04516/2020}) with the
    financial support of FCT.IP.}}

\author{Mário Pereira(\Letter)\orcidID{0000-0003-4234-5376}}
\institute{NOVA LINCS, NOVA School of Science and Technology, Lisbon, Portugal \\
\email{mjp.pereira@fct.unl.pt}}

\maketitle

\begin{abstract}
  % Despite all the tremendous recent success of deductive
  % verification, it is rarely the case that verification tools are
  % applied to programs written in functional languages. When compared
  % to the imperative world, there are only a handful of verification
  % tools that can deal with functional programs. We believe the lack
  % of pedagogical, problem-oriented documentation on how to use such
  % tools might be one of the reasons behind this apparent mismatch
  % between deductive verification and real-world functional software.

  % Functional languages have evolved from academic artifacts into
  % mature and flexible environments on which one can develop efficient,
  % real-world software. With languages from this family gaining
  % momentum within industry, one might naturally pose the question of
  % how to apply formal methods to ensure the correctness of software
  % written in a functional language.
  In this paper, we provide a comprehensive, hands-on tutorial on how
  to apply deductive verification to programs written in \ocaml. In
  particular, we show how one can use the \GOSPEL specification
  language and the \cameleer tool to conduct mostly-automated
  verification on \ocaml code. In our presentation, we focus on two
  main classes of programs: first, purely functional programs with no
  mutable state; then on imperative programs, where one can mix
  mutable state with subtle control-flow primitives, such as
  locally-defined exceptions.

\end{abstract}

\section{Introduction}
\label{sec:introduction}

Deductive software verification~\cite{filliatre11sttt} is a subject
within the larger field of formal
methods~\cite{DBLP:books/daglib/0007501}. One can define deductive
software verification as the process of expressing the correctness of
a program as a mathematical statement, then proving it. However, such
a definition does not properly highlight the connection between the
three main components in deductive verification: the \emph{logical
  specification}, which mathematically captures \emph{what} one wishes
to compute; the \emph{code}, which stands for \emph{how} one
materializes ideas as a piece of software; and, finally, a formal
proof of \emph{why} the code adheres to the given specification. This
last component can be realized via a so-called \emph{Verification
  Conditions Generator}, a mechanical tool that takes as input the
code and the specification, producing the aforementioned correctness
statement.

% Mário: eliminado por sugestão do reviewer 3
% Deductive software verification~\cite{filliatre11sttt}, in the course of the
% last two decades, has been the subject of truly remarkable progress. It is,
% nowadays, a field of intensive, active development by a large community of
% researchers. Deductive verification tools have grown into mature, usable
% frameworks that allow one to tackle the verification of realistic,
% industrial-scale projects.
% % Apesar de tudo, verificação deductiva é pouco aplicada a PF
% Yet, deductive verification has been seldom applied to programs written in a
% functional language~\cite{DBLP:conf/mpc/Regis-GianasP08}. There are currently
% only a handful of practical, easy to use tools to tackle the verification of
% functional programs. Moreover, there is a clear lack of comprehensive
% documentation (in the form of books, tutorials, courses) that clearly show how
% to use such tools to successfully tackle the verification of interesting case
% studies.
% % Estranho, pois este paradigma é particularmente bem adaptado para escrever
% % programas correctos
% This might come as a surprise, since languages from the functional paradigm
% normally provide the features to write correct code, in particular: well-defined
% semantics, simple syntax closer to mathematical definitions, state-of-the-art
% type-systems, and flexible module systems.

% Este tutorial paper quer ser uma mudança neste paradigma e mostrar que
% linguagens funcionais também merecem a atenção da comunidade.
In this tutorial paper, we focus on the deductive verification of
programs developed in the \ocaml language.
% In particular we
% show how one can use deductive verification to reason about code
% written in \ocaml.
% Apresentamos um conjunto de ferramentas para a especificação e prova formal
% de programas OCaml.
We use the \cameleer~\cite{pereira2021cameleer} tool to verify, in a
mostly-automated fashion, that an \ocaml program adheres to its
specification.
% Adoptamos uma abordagem incremental na "dificuldade" dos programas que
% queremos provar: (1) prova de programas puramente funcionais (próximos de
% definições lógicas e matemática); (2) prova de programas com mutabilidade
% controlada, isto é, sem aliasing; (3) Lógica da Separação.
%
% We follow in our presentation an incremental approach: first, we show
% how to reason about purely functional (\emph{i.e.}, with no mutable
% state) programs; second, we tackle the verification of programs
% featuring mutability, but excluding pointers manipulation; finally, we
% use a Separation Logic to prove pointer-manipulating \ocaml
% programs. The proofs in the first two steps are conducted using
% \cameleer~\cite{pereira2021cameleer}.
%
% Apesar das diferentes lógicas e ferramentas, uma linguagem de especificação
% em comum. Gospel.
One key aspect of our presentation is the use of
\GOSPEL~\cite{DBLP:conf/fm/ChargueraudFLP19}, the \emph{G}eneric
\emph{O}Caml \emph{SPE}cification \emph{L}anguage.
This is a tool-agnostic language, which serves as a common ground for
the different \ocaml verification tools and techniques. One important
feature of \GOSPEL is that specifications are written in a subset of
the OCaml language, plus quantifiers, making the adoption of formal
methods even more appealing for the working \ocaml programmer.

Throughout this tutorial, we harmoniously combine an algorithmic
discovery journey with the art of deductive software verification. We
believe verification tools and techniques are better presented through
the lens of classical data structures and algorithms. We believe this
hands-on, example-oriented approach is a more efficient and convincing
way to justify the interest in using verification tools. In order to
follow this tutorial, we only assume the reader to possess basic
knowledge of functional programming (not necessarily in \ocaml) and
some knowledge of deductive verification, at least to the level of
understanding function contracts, loop invariants, and proofs by
induction.

\mario{Dizer que este tutorial apenas assume conhecimentos básicos de
  programação OCaml e algum conhecimento de verificação dedutiva ---
  to the level of writing function contracts, loop invariants and do
  proofs by induction.}

\mario{Referir este tutorial é também um mix entre verificação e ``descoberta''
  algoritmica. Demonstrar as técnicas e ferramentas através do olhar/lente das
  estruturas de dados e algoritmos (clássicos).}

\mario{Este é também um tutorial sobre técnicas de especificação e prova, e.g.,
  permitted/complete para iteração; invariantes descentralizados.}

\mario{We believe this hands-on, example-oriented approach}

This paper is organized as follows. Sec.~\ref{sec:primer-gospel}
provides an overview of the \GOSPEL
language. Sec.~\ref{sec:camel-verif-tool} introduces the \cameleer
tool, mainly using two examples of verified \ocaml programs, the first
being a pure implementation, the other featuring mutability. In
Sec.~\ref{sec:purely-funct-progr}, we take a more in-depth dive into
the verification of functional
programs. Sec.~\ref{sec:imperative-programs} extends our class of
verified programs, incorporating some imperative traits of the \ocaml
language. We terminate with some related work
(Sec.~\ref{sec:related-work}) and closing remarks and future
perspectives (Sec.~\ref{sec:conclusion}). All the software and proofs
used in this paper are publicly available in a companion
artifact~\cite{pereira_2024_12588707}, which also complements this
paper with other case studies verified in \cameleer.

\section{A Primer on \GOSPEL}
\label{sec:primer-gospel}

\GOSPEL is a behavioral specification language for \ocaml code. It is
a contract-based, statically typed language, with a formal semantics
defined by means of translation into Separation
Logic~\cite{10.1145/3408998,10.5555/645683.664578}. The term
\emph{Generic} comes from the fact that \GOSPEL is not tied to any
particular tool or analysis technique. In fact, nowadays, one can use
\GOSPEL to attach specifications to \ocaml that are then analyzed
using \emph{runtime assertion checking}
techniques~\cite{DBLP:conf/rv/FilliatreP21}, or formally verified
using deductive verification
tools~\cite{pereira2021cameleer,DBLP:books/hal/Chargueraud23}. \GOSPEL
is inspired by other behavioral specification
languages~\cite{DBLP:journals/csur/HatcliffLLMP12}, such as
\textsf{JML}~\cite{DBLP:journals/sigsoft/LeavensBR06} or
\textsf{Eiffel}~\cite{DBLP:books/ph/Meyer91}. However, both
\textsf{JML} and \textsf{Eiffel} require the specification to always
be executable. This in not the case in \GOSPEL. In this tutorial
paper, we focus on the use of \GOSPEL for deductive verification.

When compared to other specification languages based on Separation
Logic, \emph{e.g.},
\textsf{VeriFast}~\cite{DBLP:conf/nfm/JacobsSPVPP11},
\textsf{Viper}~\cite{DBLP:conf/vmcai/0001SS16}, or
\textsf{Gillian}~\cite{DBLP:conf/cav/MaksimovicASG20}, \GOSPEL takes a
different design choice: permission and separation conditions are
implicitly associated with function arguments, which greatly improves
conciseness over Separation Logic. We argue this is an important
argument in favor of \GOSPEL adoption by regular \ocaml programmers,
who are not necessarily proof experts. We believe it is of crucial
importance to develop the languages and tools that bring practitioners
into formal methods.

% One important step is to provide the means for working programmers, who are not
% necessarily proof experts, to at least annotate their code with formal
% specification elements.

\begin{figure}[t]
\begin{gospel}
  type 'a t
  (*@ mutable model view: 'a list *)

  val create : unit -> 'a t
  (*@ s = create ()
        ensures s.view = [] *)

  val is_empty : 'a t -> bool
  (*@ b = is_empty s
        ensures b <-> s.view = [] *)

  val push : 'a -> 'a t -> unit
  (*@ push x s
        modifies s.view
        ensures  s.view = x :: old s.view *)

  val pop : 'a t -> 'a
  (*@ v = pop s
        requires s.view <> []
        modifies s.view
        ensures  v :: s.view = old s.view *)
\end{gospel}
\caption{\GOSPEL-annotated Stack Interface.}
\label{fig:stack-gospel}
\end{figure}

\GOSPEL was initially designed as an interface behavioral
specification language. The interface shown in
Figure~\ref{fig:stack-gospel} exemplifies the use of \GOSPEL to
specify an \ocaml interface for a \emph{polymorphic stack} data
structure, independent of the underlying implementation (it could be,
\emph{e.g.}, a linked-list, a ring buffer, etc.). \GOSPEL
specification is given within comments of the form \of{(*@ ... *)}.
We start by specifying that type~\of{t} of stacks is described, at the
logical level, via a \emph{model field} named \of{view}. This field is
of type \of{'a list} (here, we use \ocaml's immutable lists) and
describes the sequence of elements contained in the data
structure. There is, however, one important aspect about the use of
\of{view}: it is declared as a \of{mutable} field, which means one
should expect in-place modifications to the stack. In other
words,~\of{t} represents an imperative data structure.

When it comes to attaching specification to functions, the first line
in the \GOSPEL comments names function arguments and its return
value. To describe the behavior of a function, we mainly use three
clauses: \of{requires}, to introduce a \pre; \of{ensures}, to
introduce a \post; and \of{modifies}, which enumerates all the mutable
fields changed during the call to a function. For instance, functions
\of{create} and \of{is_empty} are simply annotated with \post{s}
stating, respectively, that a fresh stack is created with no elements
and, conversely, a stack is empty if it does not contain any
element. Finally, functions \of{push}, \of{pop}, and \of{transfer}
modify the contents of a stack via side-effects. The former inserts a
new element to the top of the \of{view} model; the latter removes the
top element, assuming as a \pre that the stack is not empty. The
term %
\of{old s.view} represents the pre-state of model \of{view},
\emph{i.e.}, the value of this field at the moment the function is
called.

% As previously mentioned, the semantics of each \GOSPEL specification triple and
% model fields is given in terms of a translation into Separation Logic. Since
% type~\of{t} features only a single model field, it carries an implicit
% \emph{representation predicate}~\coqinline{S} of type~%
% \coqinline{loc -> 'a list -> heap -> Prop},~%
% where~\of{loc} stands for the type of pointers in Separation Logic. The
% assertion \coqinline{S p L} captures the piece of state and invariants involved
% in the memory representation of a mutable stack at address \coqinline{p} with
% contents \coqinline{L}. The translation from \GOSPEL to Separation Logic, for
% stack operations, is given below:
% %
% \begin{lstcoq}
%   { \[True] } create () { fun s. \exists L. (S s L) \* \[L = nil] }
%   { S s L } is_empty s { fun b. (S s L) \* \[b = true <-> L = nil] }
%   { (S s L) } push x s { fun _. \exists L'. (S s L') \* \[L' = v::L] }
%   { (S q L) \* \[L <> nil] } pop s { fun v. \exists L'. (S s L') \* \[L = L' ++v::nil] }
% \end{lstcoq}
% %
% Observe, in particular, how one must explicitly assert the validity of
% representation predicates in pre- and \post{s}. In \GOSPEL, representation
% predicates are implicitly attached to function arguments and re-established in
% \post{s}.

Throughout this tutorial, we will use and discover \GOSPEL features
via \ocaml examples formally verified using the \cameleer
tool. However, we are not able to cover here all the relevant aspects
of the language in full detail. For a more in-depth presentation of
\GOSPEL, we refer the reader to the original
paper~\cite{DBLP:conf/fm/ChargueraudFLP19} and the user
manual\footnote{\url{https://ocaml-gospel.github.io/gospel/}}.

\section{The \cameleer Verification Tool}
\label{sec:camel-verif-tool}

\mario{É logo aqui no início que devo explicar a arquitectura do \cameleer ---
  figura do paper CAV.}

\begin{figure}[t]
  \centering
  \includegraphics[width=\linewidth]{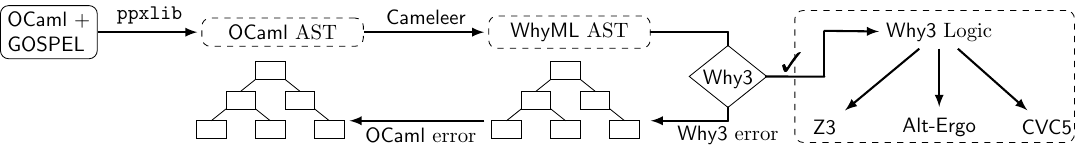}
  \caption{\cameleer Architecture and verification pipeline, taken
    from~\cite{pereira2021cameleer}.}
  \label{fig:cameleer-overview}
\end{figure}

Up until recently, programmers would face a difficult choice if they
wished to produce verified \ocaml code: either conduct automated
proof, but entirely re-implement their code-bases in a proof-aware
language, and then rely on code extraction; or verify actual \ocaml
code, but using an interactive proof assistant, with the burden of
manual proofs. \cameleer offers a compromise between the two
approaches: it is a tool for the deductive verification of
\ocaml-written programs, with a clear focus on proof automation. It
aims to provide an easy to use framework for the specification and
verification of \ocaml code.

Fig.~\ref{fig:cameleer-overview} presents the verification pipeline of
the tool. It takes as input an \ocaml implementation file, annotated
using \GOSPEL, and translates it into an equivalent \whyml program,
the programming and specification language of the \why
framework~\cite{DBLP:conf/esop/FilliatreP13}. \why is a tool-set for
the deductive verification of software, oriented towards automated
proof. A distinctive feature of \why is that it can interface with
different off-the-shelf SMT solvers, which greatly increases proof
automation.

With \cameleer, we put forward the vision of the \emph{specifying
  programmer}: those who write the code, should also be able to
specify it. But, we want to push this vision even further: those who
write the code, should be able to specify it \emph{and formally verify
  it}. Leveraging on the proof automation and tool-set offered by
\why, we believe \cameleer is a good candidate to fill this need in
the working \ocaml programmers community.

In this section, we introduce \cameleer via two examples of
mechanically verified algorithms implemented in \ocaml and specified
in \GOSPEL. The first one is a traditional merge operation over sorted
lists. The second one is a linear-search operation on arrays, hence an
imperative implementation, featuring some interesting constructions
from the \ocaml language. The two examples have a common point: both
are written using functors, showing how \cameleer proofs can scale up
to the level of modular algorithms and data structures.

\mario{Referir durante esta primeira parte da secção que o \cameleer ships with a
  parte da biblioteca standard \ocaml especificada em \GOSPEL.}

\mario{Nesta secção: um exemplo puro --- \of{merge} e apresentar o tipo
  \of{PRE_ORD}; um exemplo imperativo --- \of{find} e apresentar o tipo \of{EQUAL}.}

\subsection{A Simple Functional Program -- the Merge Routine}
\label{sec:simple-funct-progr}

\mario{Fazer notar aqui que só estamos interessados em mostrar \of{merge}
  devolve uma lista ordenada, não mostramos que se trata de uma permutação dos
  seus argumentos --- dizer isto logo no início da secção. --- trata-se de um
  exemplo motivador.}

\begin{figure}[t]
     \begin{gospel}
module type PRE_ORD = sig
  type t

  (*@ predicate le (x : t) (y : t) *)

  (*@ axiom reflexive : forall x. le x x *)
  (*@ axiom total     : forall x y. le x y \/ le y x *)
  (*@ axiom transitive: forall x y z. le x y -> le y z -> le x z *)

  val leq : t -> t -> bool
  (*@ b = leq x y
        ensures b <-> le x y *)
end
\end{gospel}
\caption{Type Equipped With a Total Preorder Relation.}
\label{fig:order}
\end{figure}

\subsubsection{Modular Definitions, Using Functors.}
\label{sec:modul-defin-using}

The \ocaml module system, in particular \emph{functors}, offers
flexible mechanisms to derive implementations that are agnostic to the
actual representation of manipulated data types. Functors stand for
modules that take other modules as parameters, similar to
\textsf{Scala} traits. As an introductory example on the use of
functors, consider the following implementation of a \of{max} function
within functor \of{Max}:
\begin{gospel}
  module Max (E : PRE_ORD) = struct
    let max x y =
      if E.leq x y then y
      else x
  end
\end{gospel}
The signature type \of{PRE_ORD} is given in Fig.~\ref{fig:order}. It
introduces a type~\of{t} together with a function \of{leq}, which
establishes a total preorder on values of type~\of{t}. The \GOSPEL
specification in this module type introduces a predicate~\of{le},
which we assume it respects the three laws of a preorder:
\emph{reflexivity}, \emph{totality}, and \emph{transitivity}. Such
laws are encoded as \emph{axioms}, which stand for logical assumptions
upon which one relies without actually providing them
correct. Finally, the \post of program function \of{leq} states this
is a decidable implementation of predicate \of{le}.

For the above implementation of \of{max}, we use the \of{leq} function
provided in the functor argument~\of{E}, to check whether function
argument~\of{x} is less or equal to~\of{y}. This comparison is made
with respect to the preorder relation induced by~\of{E}. % To use the
% \of{max} definition, it is up to the programmer to instantiate module
% \of{Max} with a concrete definition for type~\of{t} and
% function~\of{leq}. For instance, the following
% %
% \begin{gospel}
%   module MInt = Max(struct type t = int val leq = (<=) end)
% \end{gospel}
% %
% instantiates the functor for integer values, where the preorder
% relation is represented by the \ocaml less-or-equal infix
% operator~\of{(<=)}. One can then write \of{MInt.max} to use the above
% \of{max} definition, specialized for integers.
%
We can provide this implementation with suitable \GOSPEL
specification. This is as follows\footnote{In \cameleer, function
  specification is introduced after function definition.}:
\begin{gospel}
  let max x y = ...
  (*@ r = max x y
        ensures E.le x r /\ E.le y r
        ensures r = x \/ r = y
\end{gospel}
The first clause in the postcondition states that both~\of{x}
and~\of{y} must be smaller than or equal to the returned value~\of{r},
with respect to the preorder induced by predicate \of{E.le}. The
second clause states that~\of{r} must be either equal~\of{x}
or~\of{y}, where \of{(=)} stands for polymorphic, potentially
undecidable, logical equality.

\subsubsection{\ocaml Implementation.}
\label{sec:ocaml-implementation-1}

The \emph{merge sort} algorithm gets its name from its main step:
merging the elements of two sorted lists, producing a third sorted
list.
% Mário: eliminado por sugestão do reviewer 3
% In a functional programming setting, writing this \of{merge}
% function is quite natural and intuitive.
%
Here, our goal is to provide a \of{merge} implementation
\emph{independently} of the type of list elements. To do so, we
introduce the following functor \of{Merge}:
\begin{gospel}
  module Merge (E : PRE_ORD) = struct
    type elt = E.t

    let rec merge_aux acc l1 l2 =
      match (l1, l2) with
      | [], l | l, [] -> List.rev_append acc l
      | x :: xs, y :: ys ->
          if E.leq x y then merge_aux (x :: acc) xs l2
          else merge_aux (y :: acc) l1 ys

    let merge l1 l2 = merge_aux [] l1 l2
  end
\end{gospel}
The above \of{merge} definition is an efficient implementation of a
merge routine, since it makes use of the tail-recursive, auxiliary
function \of{merge_aux}. This function merges lists~\of{l1}
and~\of{l2} into the accumulator~\of{acc}. Since every new element is
inserted to the head of \of{acc}, then in the base cases one must
first reverse it and then concatenate the result with~\of{l}, the
suffix list of elements (either from~\of{l1} or~\of{l2}) that remains
to be enumerated. The \ocaml standard library function \of{rev_append}
efficiently implements this ``reverse then concatenate'' process.
\mario{não sei se esta frase sobre a biblioteca standard \ocaml em
  \cameleer faz muito sentido aqui.}
%
% \cameleer ships with a subset of the \ocaml standard library
% specified using \GOSPEL, hence one is able to use and reason about
% functions from common modules, such as \of{List} or
% \of{Array}.
%
Finally, the main \of{merge} function calls \of{merge_aux} with the
empty list as the initial value for the accumulator.

\subsubsection{\GOSPEL Specification.}
\label{sec:gospel-specification}

Let us now describe the specification of the \of{merge_aux} and
\of{merge} functions. In order to specify that \of{merge} always
returns a sorted list, we must first introduce what it means for a
list to be sorted. We introduce the following \GOSPEL predicate:
\begin{gospel}
  (*@ predicate rec sorted_list (l : elt list) =
        match l with
        | [] | _ :: [] -> true
        | x :: y :: r -> E.le x y && sorted_list (y :: r) *)
  (*@ variant l *)
\end{gospel}
The empty or singleton lists of integers are always sorted. If the
list has at least two elements, then the first must less than or equal
to the second one and the suffix list \of{y :: r} must also be a
sorted list. Since \of{sorted_list} is to be used within
specifications, it must be a total (\emph{i.e.}, terminating)
function. We supply the variant \of{l}, which represents a
\emph{termination measure} for every recursive call to the
\of{sorted_list} predicate. A variant represents a quantity that
always strictly decreases at every recursive call. In this case, we
state that the argument of every recursive call is structurally
smaller than the value of~\of{l} at the entry point.

Note that the elements of the argument~\of{l} are of type
\of{integer}, the \GOSPEL type for mathematical integers. When
applying this function to a list of values of \ocaml \of{int} type
(63-bit machine integers), the \GOSPEL and \cameleer tool-chains will
apply a conversion mechanism from machine integers into their
equivalent mathematical representation.

Using \of{sorted_list} predicate, we attach the following \GOSPEL
specification to the \of{merge_aux} function:
\begin{gospel}
  let rec merge_aux acc l1 l2 = ...
  (*@ r = merge_aux acc l1 l2
        requires sorted_list (List.rev acc)
        requires sorted_list l1 && sorted_list l2
        requires forall x y.
          List.mem x acc -> List.mem y l1 -> E.le x y
        requires forall x y.
          List.mem x acc -> List.mem y l2 -> E.le x y
        ensures  sorted_list r
        variant  l1, l2 *)
\end{gospel}
The \pre reads as follows: the \of{acc} list is sorted in reverse
order, while~\of{l1} and~\of{l2} are sorted in natural order; every
element from~\of{acc} must be less or equal to any element from
either~\of{l1} or~\of{l2}. Finally, the \post simply asserts the
returned list~\of{r} is sorted and we prove termination using the
lexicographic order on the pair~%
\of{l1, l2}. In other words, if in a recursive call~\of{l1}
structurally decreases, then the whole variant decreases; otherwise,
when~\of{l1} does not decrease, then it must be the case that~\of{l2}
decreases. \cameleer ships with a subset of the \ocaml standard
library specified using \GOSPEL, hence one is able to use and reason
about functions such as \of{List.mem} or \of{List.rev}.

Finally, the specification of \of{merge} is as follows:
\begin{gospel}
  let merge l1 l2 = merge_aux [] l1 l2
  (*@ r = merge l1 l2
        requires sorted_list l1 && sorted_list l2
        ensures  sorted_list r *)
\end{gospel}
If both~\of{l1} and~\of{l2} are sorted lists, then the call %
\of{merge l1 l2} always produces a sorted list.

\subsubsection{\cameleer Proof.}
\label{sec:cameleer-proof-merge}

\begin{figure}[t]
  \centering
  \includegraphics[width=\linewidth]{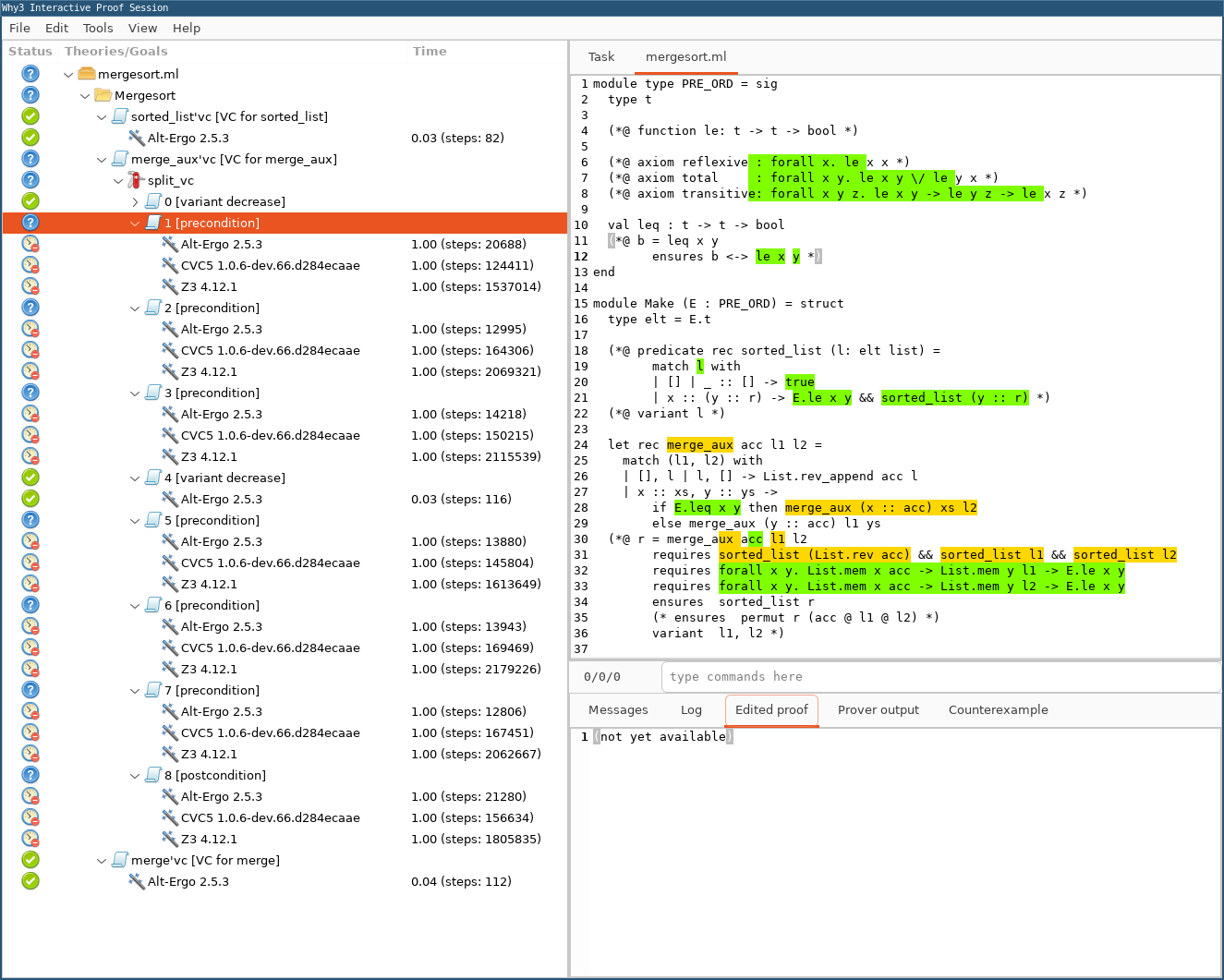}
  \caption{\why Proof Session for the Merge Sort Routine.}
  \label{fig:why3-merge}
\end{figure}

Assuming the \ocaml implementation and \GOSPEL specification of the
merge routine are contained in file \texttt{merge.ml}, one can start
the verification process by typing the command \texttt{cameleer
  merge.ml}. This launches the interactive \why graphical user
interface~\cite{dailler2018}, as depicted in
Fig.~\ref{fig:why3-merge}. On left-hand side, the \why IDE features a
node for the generated verification conditions (VCs) of each top-level
definition, together with any proof attempt on such VCs. Calling a SMT
solver to prove a generated VC can be done by right-clicking a node
and selecting the desired solver. A green button on the left of a node
means that a solver was able to discharge the corresponding VC. This
is the case of the \of{sorted_list} predicate and the \of{merge}
function, for which \textsf{Alt-Ergo} is able to prove that both
adhere to their specification, in less than a second. The proof time
is shown on the right of the solver name, together with the number of
conducted proof steps. One can also apply proof transformations on
nodes, for instance to split a larger VC into its conjunctive
clauses. This is done for the \of{merge_aux} definition. After split,
one can focus on a specific part of the verification, such as proving
a precondition, variant decrease, or postcondition. Moreover, each of
these individual formulae is smaller and less involved than the
original VC, hence more likely to be automatically discharged by an
SMT solver.

On the right-hand side of the \why IDE, one can inspect the code under
verification. On one hand, the green labels stand for the assumptions
(flow of the execution and parts of the specification) made at some
point of the verification process. On the other hand, the yellow
labels mark what one is actually trying to verify. In this case, we
are trying to discharge the first precondition of the \of{merge_aux}
function, at the recursive call %
\of{merge_aux (x :: acc) xs l2}. We shall explain the \texttt{Task}
tab in Sec.~\ref{sec:warm-up-searching}.

As shown in Fig.~\ref{fig:why3-merge}, we fail to prove that function
\of{merge_aux} adheres to its \GOSPEL specification. We are not able
to verify every individual VC for this function, after split. The
three used solvers, \textsf{Alt-Ergo}, \textsf{CVC5}, and \textsf{Z3},
all time-out after 1 second for every condition except that the
supplied variant measure decreases. One could attempt to provide more
time to each solver, to allow these tools to conduct more proof steps,
hopefully leading to more VCS being discharged. However, in this case,
we are actually missing some \emph{auxiliary lemmas} about the
\of{sorted_list} predicate. A lemma in \GOSPEL represents a property
that, once stated, can be explored by SMTs to discharge other
VCs. However, contrarily to axioms, lemmas are not assumed: these must
be proved correct at definition time.

To close the verification of the \of{merge_aux} function, we need two
auxiliary lemmas: first, one that states we can insert a new
value~\of{x} as the head of a sorted list~\of{l} if and only if~\of{x}
is less than or equal to every element in~\of{l}. We introduce such a
lemma in \GOSPEL as follows:
\begin{gospel}
  (*@ lemma sorted_mem: forall x l.
        (forall y. List.mem y l -> E.le x y) /\ sorted_list l <->
        sorted_list (x :: l) *)
\end{gospel}
Second, the concatenation \of{l1 @ l2} of sorted lists~\of{l1} and~\of{l2} is a
sorted list if and only if all the elements in~\of{l1} are less or equal than
all the elements of~\of{l2}. We provide the following \GOSPEL lemma:
\begin{gospel}
  (*@ lemma sorted_append: forall l1 l2.
        (sorted_list l1 && sorted_list l2 &&
         (forall x y. List.mem x l1 -> List.mem y l2 -> E.le x y))
    <-> sorted_list (l1 ++ l2) *)
\end{gospel}
Using the given lemmas, the correctness proof for \of{merge_aux} now
succeeds. \textsf{Alt-Ergo} is able to explore these auxiliary
definitions to discharge all the remaining VCs. As for the proofs of
the lemmas themselves, both require proofs by induction. We can
conduct such proofs inside the \why IDE, using a dedicated
transformation for induction over algebraic types (in this case
lists). For a more complete presentation on how to use the \why IDE,
including on how to apply different interactive proof transformations,
we refer the reader to the framework user's
manual~\cite{bobot2011why3}.

% conduct such proofs using the \why
% Interactive IDE~\cite{dailler2018}, which provides lightweight
% interactive tactics, like for proofs by induction, without sacrificing
% the automation of SMT-based proofs.

% For the \of{merge} implementation, the \cameleer-\why pipeline
% generates a total of 28 verification conditions (VC): one to prove
% \of{sorted_list} terminates; 6 and 8 for lemmas \of{sorted_mem} and
% \of{sorted_append}, respectively, after applying the \of{induction}
% tactic; 9 for \of{merge_aux}; and 4 for \of{merge}. Each VC is
% discharged in less than one tenth of a second, using only the
% \textsf{Alt-Ergo} SMT solver.

\subsection{Searching an Element Within an Array}
\label{sec:warm-up-searching}

\begin{figure}[t]
    \begin{gospel}
  module type EQUAL = sig
    type t

    val eq : t -> t -> bool
    (*@ b = eq x y
          ensures b <-> x = y *)
  end
\end{gospel}
\caption{Type Equipped With an Equality Relation.}
\label{fig:equal}
\end{figure}

\subsubsection{\ocaml Implementation.}
\label{sec:ocaml-implementation-2}

We now make a shift from the purely functional world to present some
\ocaml imperative features, and showcase how one can use \cameleer to
reason about such features.
Consider the following modular implementation of a function that
performs a linear search in an array:
\begin{gospel}
  module Find (E : EQUAL) = struct
    let find x a =
      let exception Found of int in
      try
        for i = 0 to Array.length a - 1 do
          if E.eq a.(i) x then raise (Found i)
        done;
        raise Not_found
      with Found i -> i
  end
\end{gospel}
Fig.~\ref{fig:equal} presents the definition of the signature type
\of{EQUAL}. It declares a function \of{eq} that decides whether two
values of type~\of{t} are \emph{logically equal}. This is exactly what
is stated in the postcondition clause.

The definition of the \of{find} function presents some interesting
\ocaml imperative traits. On one hand, the use of a \of{for} loop to
scan the array~\of{a}; on the other hand, the declaration and use of
the \emph{local exception} \of{Found}. The latter is used to signal
that the search succeeded, carrying the index of~\of{x}
within~\of{a}. The use of local exceptions in \ocaml is a convenient
way to simulate the behavior of a \of{return} statement, commonly
found in other languages. In fact, the whole loop is surrounded with a
\of{try..with} block that ensures exception \of{Found} is always
caught. Finally, to signal that~\of{x} does not occur in~\of{a}, we
use the \of{Not_found} exception from the \ocaml standard library. It
is worth noting that we purposely let such an exception escape the
scope of \of{find}.

\subsubsection{\cameleer Proof.}
\label{sec:cameleer-proof-2}

% The complete \ocaml implementation and \GOSPEL specification for the \of{find}
% example can be found in Appendix~\ref{sec:find-function}.

In order to prove the correctness of function \of{find}, one must
supply a \emph{loop invariant}. This is done in \cameleer as follows:
\begin{gospel}
        for i = 0 to Array.length a - 1 do
          (*@ invariant forall j. 0 <= j < i -> a.(j) <> x *)
\end{gospel}
This invariant simply asserts that while in the loop, we know for
sure~\of{x} does not occur in the prefix of~\of{a} that we have
already scanned. The infix operator \of{(<>)} stands for logical
inequality. As for the equality operator, \of{(=)}, inequality is
expressed using the same syntax in \ocaml and in \GOSPEL and both are
built-in symbols of the \GOSPEL language. We recall that, except for
quantifiers and logical connectives, \GOSPEL terms are a written in a
subset of the \ocaml language.

Now, we focus on providing a specification contract for function
\of{find}. But first, it is crucial to distinguish the possible
outcome behaviors of this function. On one hand, it returns normally
whenever exception \of{Found} is raised; on the other hand, it raises
the \of{Not_found} exception to abort execution. For the former, we
shall establish a \emph{regular \post}. For the latter, we shall
introduce what is called an \emph{exceptional \post}. We attach the
following \GOSPEL annotations to \of{find}:
\begin{gospel}
    let find x a =
      ...
    (*@ i = find x a
          ensures a.(i) = x
          raises Not_found -> forall i. 0 <= i < Array.length a ->
            a.(i) <> x *)
\end{gospel}
The \of{ensures} clause is checked when \of{find} indeed returns an
integer~\of{i}, representing the (first) index of~\of{x}
in~\of{a}. The \of{raises} clause states the logical property that
holds when \of{Not_found} is raised. This stands for the case when we
have scanned all the array~\of{a}, finding no occurrence of~\of{x}. We
restrict the range of values that the universally quantified
variable~\of{i} can take, since \GOSPEL establishes that undefined
array indices are arbitrary values, not necessarily different
from~\of{x}.

For this program, \cameleer generates 6 VCs, after splitting the
formula generated for the \of{find} function. All are immediately
discharged by \textsf{Alt-Ergo}.

\subsubsection{Providing an Incorrect Loop Invariant.}
\label{sec:providing-wrong-loop}

\begin{figure}[t]
  \centering
  \includegraphics[width=\linewidth]{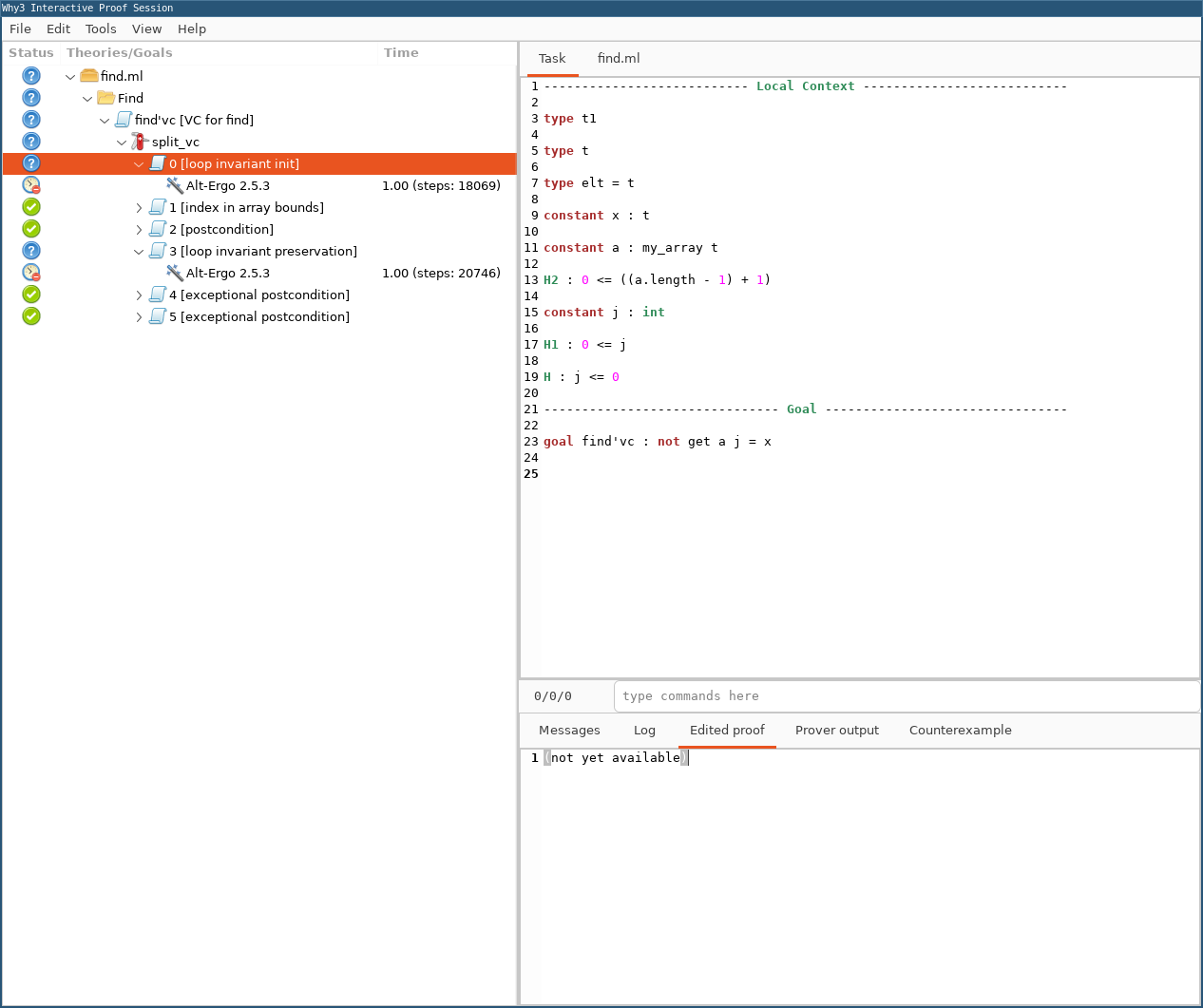}
  \caption{Proof Task for an Incorrect Loop Invariant Initialization.}
  \label{fig:why3-mjrty}
\end{figure}

Let us take a step back in the verification process of the \of{find}
implementation. Imagine a scenario where one would have, incorrectly,
supplied the following loop invariant:
\begin{gospel}
        for i = 0 to Array.length a - 1 do
          (*@ invariant forall j. 0 <= j <= i -> a.(j) <> x *)
\end{gospel}
The only difference, when compared with the previously presented
invariant, is that now the value of the universally quantified
variable~\of{j} can be equal to~\of{i}, the loop
index. Fig.~\ref{fig:why3-mjrty} shows that by feeding the new
invariant to the \cameleer-\why pipeline, one is still able to prove
the postcondition of \of{find} holds, but not the \emph{invariant
  initialization} (\emph{i.e.}, the invariant holds before the first
iteration), neither \emph{invariant preservation} (\emph{i.e.},
assuming the invariant holds before an arbitrary iteration, it still
holds after that iterations completes). To debug a failed proof
attempt, the \why IDE allows the user to inspect the
\emph{task}~\cite{boogiewhy3}, a representation of the formula that is
sent to solvers. Under tab \texttt{Task}, such formula is displayed as
a \of{goal} (\emph{i.e.}, what one is actually trying to prove) and
the proof context (\emph{i.e.}, the hypotheses) above the dashed line.

Fig.~\ref{fig:why3-mjrty} depicts the task for the loop invariant
initialization. After reading the task, we can conclude that
hypotheses \of{H1} and \of{H2} imply that~\of{j} is equal to~0. Hence,
we are trying to prove a goal that asserts the actual first element in
the array~\of{a} is not~\of{x}. There is nothing in our proof context
that allows us to prove such a statement. This is an indication that
either our context is not enough to discharge the goal (\emph{e.g.},
the specification is incomplete), or rather there is an actual bug in
the specification or implementation. In this case, however, we already
know the answer: changing the loop invariant to \of{j < i} would
generate a task, for loop invariant initialization, with hypotheses
\of{0 <= j} and \of{j < 0}, hence the goal would hold vacuously. It is
worth noting that tasks are written in the logical fragment of the
\whyml language. This is the only point in the verification process
that a \cameleer user must read a formula that is not written in
\GOSPEL. However, since \GOSPEL and \whyml are syntactically very
similar, we believe someone familiar with \GOSPEL is able to read and
understand a \why task.

\section{Purely Functional Programming}
\label{sec:purely-funct-progr}

We define \emph{purely functional programming} as the approach to
write a program where no mutable state is involved. Such a program is
normally a collection of (recursive) functions that are composed with
each other to perform some computation. The program also only employs
data structures that do not require any memory manipulation, such as
lists or trees. This style of programming is at the very essence of the
\ocaml language.

In this section, we use the \cameleer tool to conduct formal
verification of a purely functional program that operates on trees. We
present the \ocaml implementation of such a program, annotated with
suitable \GOSPEL specifications. We interleave the presentation of
code listings with explanations of the given implementation and
specification. Hence, to ease readability, the example chunks that
belong together are given running line numbers.

\subsection{Same Fringe -- Comparing Two Binary Trees}
\label{sec:same-fringe-comp}

Let us consider the following, very classic, programming challenge:
\begin{quote}
  \textit{Write a function that, given two binary trees, decides whether the two
    trees present the same sequence of elements when traversed inorder.}
\end{quote}
This is known as the \emph{same fringe} problem.\mario{quote?} One
possible solution to this problem is as follows:
\begin{enumerate}
\item perform an inorder traversal on both trees, building the list of elements
  enumerated during such traversals;
\item compare, recursively, whether the two lists contain the same elements.
\end{enumerate}
This is, however, a very naive approach, since it always builds the auxiliary
lists for all the elements of both trees. Imagine the following scenario:
\begin{center}
\begin{tikzpicture}[scale=.5,sibling distance=8em, level distance=4em,
  every node/.style = {shape=circle, scale=1, draw,
    minimum size = 1.5em, thick},
  triangle/.style={isosceles triangle,draw=,thick,shape border
    rotate=90,isosceles triangle stretches=true, minimum height=1mm,minimum
    width=5mm,inner sep=0,yshift={-10mm}},
  very large triangle/.style={triangle,minimum width = 33mm,minimum
    height=1mm,yshift={-12mm}}]
  \node (root) {} [child anchor = north]
  child {node (x) {\texttt{x}}}
  child {node[very large triangle] {huge sub-tree}};

  \node [right = 6.5cm of root] (root2) {} [child anchor = north]
  child {node (y) {\texttt{y}}}
  child {node[very large triangle] {huge sub-tree}};
\end{tikzpicture}
\end{center}
The two trees differ in the leftmost element, as we have~\of{x} for
the left-hand side tree, and~\of{y} for the right-hand side
tree. Following the solution proposed above, we would unnecessarily
build two (huge) sequences. Note, however, that this implementation is
easier to check for correctness than more efficient ones. Hence, this
list-based approach can be used as a good specification for the
solution we describe in the remaining of the section.

We propose to explore an approach that allows one to enumerate the
elements of each tree, step-by-step. In such a way, we can stop as
soon as two distinct elements are enumerated. If we complete the
iteration process on both trees, then it must be the case the trees
contain the same elements.
This algorithm is implemented in \ocaml and specified using \GOSPEL as
follows. First, we use the \of{EQUAL} signature from
Fig.~\ref{fig:equal} to build a functor that implements \emph{same
  fringe}. We start by defining the type of binary trees with elements
of type \of{E.t}, as follows:
\begin{gospelnumbers}
  module Make (E : EQUAL) = struct
    type tree = Empty | Node of tree * E.t * tree
\end{gospelnumbers}
A \texttt{tree} is either \of{Empty} or a \of{Node} formed of two sub-trees and
a root of type~\of{E.t}.

Now, we define a logical function that implements an inorder traversal on a
binary tree, returning the list of enumerated elements:
\lstset{firstnumber=3}
\begin{gospelnumbers}
    (*@ function elements (t : tree) : E.t list =
          match t with
          | Empty -> []
          | Node (l, x, r) -> (elements l) @ (x :: elements r) *)
\end{gospelnumbers}
The traversal is implemented by: (i) traversing the whole left
sub-tree; (ii) concatenating the resulting sequence of elements (the
\of{(@)} operator represents list concatenation in \ocaml) to \of{x}
(the root) and the sequence of elements issued from the right sub-tree
traversal. This is exactly the naive approach previously described. We
shall only use the function \of{elements} for specification purposes.

In order to implement the step-by-step enumeration of elements in a
tree, we use an explicit data representation of the call stack of the
program that would construct the inoder list, so we can interrupt the
traversal as soon as required. Such a representation is inspired by
the \emph{zipper}~\cite{zipper} structure.%
\mario{dizer mais alguma coisa sobre o zipper?}  A zipper can be used
as an efficient cursor into data structures, allowing one to
arbitrarily traverse the structure, as well as to perform efficient
local modifications (\emph{e.g.}, insertions or deletions) without the
overhead of rebuilding the whole structure for each modification. In
the case of the \emph{same fringe} problem, we always traverse a tree
towards its leftmost element, without performing any modifications to
the structure. Hence, we specialize the zipper data type as follows:
\lstset{firstnumber=7}
\begin{gospelnumbers}
    type zipper = (E.t * tree) list
\end{gospelnumbers}
A value of type \of{zipper} is a list, where each element is a pair
composed of a tree element and the corresponding right sub-tree. We
build such a list bottom-up, which represents the left spine of the
tree that is still to be traversed. Together with the \of{zipper} data
type, we introduce the following logical function to convert from a
\of{zipper} to a list:
\lstset{firstnumber=8}
\begin{gospelnumbers}
    (*@ function enum_elements (e : zipper) : E.t list =
          match e with
          | [] -> []
          | (x, r) :: e -> x :: (elements r @ enum_elements e) *)
\end{gospelnumbers}
\mario{Dizer que de facto esta função representa de forma lógica a sequência de
  todos os elementos enumerados pelo zipper. Isto seria código OCaml válido, mas
  apenas precisamos de o manter na lógica já que seria desnecessário e pouco
  eficiente fazê-lo no ``mundo real''.}

\mario{Dizer alguma coisa sobre a terminação destas funções lógicas?}

From a specification point of view, we have everything we need to
tackle our verified implementation of the \emph{same fringe}
problem. We start by defining how to create a zipper from a tree. This
is done as follows:
\lstset{firstnumber=12}
\begin{gospelnumbers}
    let rec mk_zipper (t : tree) (e : zipper) =
      match t with
      | Empty -> e
      | Node (l, x, r) -> mk_zipper l ((x, r) :: e)
   (*@ r = mk_zipper t e
         variant t
         ensures enum_elements r = elements t @ enum_elements e *)
\end{gospelnumbers}
%
% We implement \of{mk_zipper} as a \emph{tail-recursive}
% function. This means that we build the zipper using an accumulator,
% argument~\of{e}, functionally updating this value at each recursive
% call.
The specification of \of{mk_zipper} states that this is a terminating
function, with argument~\of{t} structurally decreasing at each
recursive call. The functional behavior of this function is captured
in the postcondition, where we state the sequence of elements of the
resulting zipper is the same as the inorder sequence of elements from
tree~\of{t}, plus the elements of the accumulator~\of{e}.

% As we will observe later, such a postcondition is sufficient to show that, if
% two zippers are created using the empty list as the initial value of the
% accumulator, if

We now provide the actual, step-by-step, iteration on two zippers, as follows:
\lstset{firstnumber=19}
\begin{gospelnumbers}
    let rec eq_zipper (e1 : zipper) (e2 : zipper) =
      match (e1, e2) with
      | [], [] -> true
      | (x1, r1) :: e1, (x2, r2) :: e2 -> E.eq x1 x2 &&
          eq_zipper (mk_zipper r1 e1) (mk_zipper r2 e2)
      | _ -> false
    (*@ b = eq_num e1 e2
          variant List.length (enum_elements e1)
          ensures b <-> enum_elements e1 = enum_elements e2 *)
\end{gospelnumbers}
This implementation distinguishes three cases:
\begin{enumerate}
\item If both zippers are empty, then we are sure to have enumerated the same
  sequence of elements.
\item If both zippers still have elements, then the next elements in the
  enumeration are the heads of both lists. We compare these and proceed
  recursively, only if the \of{E.eq x1 x2} comparison holds.
\item Otherwise, if one of the zipper terminates before the other, then we are
  sure the enumerated sequences differ.
\end{enumerate}
Very simply put, the postcondition of \of{eq_zipper} states that this function
logically decides whether two zippers enumerate the same sequence of elements.

Finally, we provide the definition of the \of{same_fringe} function, which
decides whether two binary trees present the same elements. This is as simple as
\lstset{firstnumber=28}
\begin{gospelnumbers}
    let same_fringe (t1 : tree) (t2 : tree) =
      eq_zipper (mk_zipper t1 []) (mk_zipper t2 [])
    (*@ b = same_fringe t1 t2
          ensures b <-> elements t1 = elements t2 *)

  end
\end{gospelnumbers}
From the postcondition of \of{eq_zipper} one can deduce that
\of{same_fringe} will return the Boolean \of{true} if and only if the
two zippers passed as arguments represent the same sequence of
elements. Since we use the empty list as the initial accumulator value
for the creation of both zippers, then the postcondition of
\of{mk_zipper} states the created zippers enumerate the exact same
sequence of elements as those of trees~\of{t1} and~\of{t2},
respectively. Hence, the postcondition of \of{eq_zipper} always holds.

Feeding our \emph{same fringe} implementation to \cameleer generates
11 VCs, after splitting each top-level definition. These are
discharged in roughly 1 second using a combination of the
\textsf{Alt-Ergo}, \textsf{Z3}, and \textsf{CVC5} SMT solvers. For
reference, the complete \ocaml implementation and \GOSPEL
specification for \emph{same fringe} is given in
Appendix~\ref{sec:same-fringe-impl}.

\subsection{Summary}
\label{sec:summary}

In this section, we used the \emph{same fringe} example to motivate
and showcase the use of \cameleer for the deductive verification of
purely functional algorithms. Such functional implementations are
closer to mathematical definitions, hence are normally easier to
reason about. Other than the example presented in this section,
\cameleer has been successfully used to prove the correctness of
real-world functional \ocaml data structures. We highlight the
\texttt{Set} module from the \ocaml standard library, and the Leftist
Heap implementation issued from the widely used
\texttt{ocaml-containers}
library\footnote{\url{https://github.com/c-cube/ocaml-containers}}. Even
if not presented in the body of this document, all such case studies
are included in the companion artifact.

\mario{O que aprendemos nesta secção. Que outras provas interessantes já foram
  feitas em \cameleer com esta abordagem.}

\mario{Exemplo grande: leftist heaps. Referir que é um exemplo retirado
  diretamente da bilioteca OCaml Containers.}

\section{Imperative Programs}
\label{sec:imperative-programs}

One important aspect of the \ocaml language is the fact that it is a
multi-paradigm language, combining functional with imperative and
object-oriented programming. In this section, we use \cameleer to
conduct the formal verification of an \ocaml program that implements
an historical algorithm using imperative features, namely loops,
mutable references, and exceptions. As in
Sec.~\ref{sec:purely-funct-progr}, the main example code listings are
presented using running line numbers.

\subsection{Boyer-Moore MJRTY Algorithm}
\label{sec:boyer-moore-majority}

Let us, once again, use an algorithmic problem as the vehicle to showcase how to
specify an \ocaml program using \GOSPEL, and how to prove it in
\cameleer. Consider the following challenge:
\begin{quote}
  \textit{Write a function that, given an array of votes, determines the
    candidate with the absolute majority, if any.}
\end{quote}
\mario{aqui talvez deva explicar melhor o que são os votos; o tipo dos votos}
The more direct solution would take the following steps:
\begin{enumerate}
\item For a total of $N$ candidates, we first allocate an array of integer
  values of length $N$ that serves as an histogram.
\item We do a first pass over the array of votes, summing up in the histogram
  the number of votes for each candidate.
\item Finally, we iterate over the histogram to check if any of the candidates
  achieves majority.
\end{enumerate}
This approach could certainly be implemented in \ocaml and proved
correct in \cameleer. It runs in $\mathcal{O}(M + N)$ time (build the
histogram, then iterate over it), where $M$ is the number of votes
. However, this approach allocates an extra memory space of $N$ cells,
\emph{i.e.}, the histogram.
Here, we adopt a different solution, due to R. Boyer and
J. Moore~\cite{Moore91}. Such a solution uses at most $2M$ comparisons
and constant extra space (other than the array of votes itself).

First, we build a functor parameterized with module type \of{EQUAL}
from Fig.~\ref{fig:equal}, so that we are able to compare candidates:
\lstset{firstnumber=1}
\begin{gospelnumbers}
  module Mjrty (E : EQUAL) = struct
    type candidate = E.t

    let mjrty a =
      let exception Found of candidate in
\end{gospelnumbers}
We begin by declaring local exception \of{Found}, which we use to
terminate the search and signal if some candidates reaches
majority. We now introduce the only extra auxiliary references we
need:
\lstset{firstnumber=6}
\begin{gospelnumbers}
      let n = Array.length a in
      let cand = ref a.(0) in
      let k = ref 0 in
\end{gospelnumbers}
The use of references \of{cand} and~\of{k} is the actual core of the
Boyer-Moore's method. \mario{Explicar melhor estas variáveis.}

We do a first traversal on the array of votes, updating \of{cand}
and~\of{k} accordingly:
\lstset{firstnumber=9}
\begin{gospelnumbers}
      try
        for i = 0 to n - 1 do
          if !k = 0 then begin
            cand := a.(i);
            k := 1 end
          else if E.eq !cand a.(i) then incr k
          else decr k
        done;
\end{gospelnumbers}
Very briefly, the loop body does the following:
\begin{itemize}
\item If the value stored in \of{k} is zero, then we change the candidate stored
  in \of{cand} to the $i$-th element of~\of{a} and update~\of{k} to
  one (lines 11 to 13);
\item If the $i$-th candidate in \of{a} is equal to the value stored in
  \of{cand}, then we increment reference \of{k} (line 14);
\item Otherwise, we decrement \of{k} (line 15).
\end{itemize}
So now, a crucial question arises: what are the invariants for this
loop that allow verification to succeed? One crucial part of the
process is to reason about ``\textit{the number of votes for a certain
  candidate in a slice of the array}''. To be able to express such
notion in the specification, we declare the following \GOSPEL
function:
\begin{gospel}
  (*@ function numof_eq (a : 'a array) (v : 'a) (l u: integer) :
        integer *)
\end{gospel}
This function represents the number of elements from~\of{a}, within
range $[\mathtt{l}; \mathtt{u})$, that are equal to value~\of{v}. For
now, we focus on using \of{numof_eq} to establish the loop
invariants. Later in this section, we provide a proper definition for
this function.

Let us a conduct a step-by-step analysis on how references \of{cand}
and \of{k} are used together within the loop, as to derive the loop
invariants. In fact, the invariants we present here are already given
in Boyer and Moore's original work~\cite{Moore91}, and we adapt those
into \GOSPEL:
\begin{enumerate}
\item The number of votes for candidate \of{cand}, within the array prefix
  already scanned, is at least~\of{k}. We write this down in \GOSPEL
  as follows:
\begin{gospel}
(*@ invariant 0 <= !k <= numof_eq a !cand 0 i
\end{gospel}
This invariant is maintained by the case analysis implemented from
line 11 to 15 in the code snippet above: we update the value of
\of{cand} whenever \of{!k} reaches zero, hence \of{!k} never stores a
negative number; we update the value stored in~\of{k}, without
changing the candidate, hence~\of{k} is kept as a lower bound for the
actual number of occurrences of \of{!cand} in the scanned part of the
array. In other words, we are sure we only decrement~\of{k} after a
sufficient number of increments occurred (except if we decrement after
reference \of{cand} is updated, but in this case~\of{k} is update to
one, hence an implicit increment has also occurred).

\item The actual number of votes for \of{cand} minus the value of~\of{k} cannot
  exceed \of{(i - !k)/2}:
\begin{gospel}
    invariant 2 * (numof_eq a !cand 0 i - !k) <= i - !k
\end{gospel}
Instead of a division, we write such an invariant using an equivalent
multiplication. The reason is somehow low-level and tied to the
upcoming verification effort: SMT solvers are known to handle
multiplication better than division.

\item For every candidate \of{c} other than \of{cand}, the number of
  votes for~\of{c}, within the scanned prefix of the array, cannot
  exceed~%
  \of{(i - !k)/2}:
\begin{gospel}
    invariant forall c. c <> !cand ->
                2 * numof_eq a c 0 i <= i - !k *)
\end{gospel}
This invariant implies that no other candidate, other than the one
stored in \of{cand}, can have the majority of votes in the slice of
the array already processed by the algorithm. Once again, we use a
multiplication by two to avoid the division.

\end{enumerate}
The last invariant actually allows one to deduce a crucial property:
after scanning all the array, \of{cand} is the only candidate that can
effectively reach majority.

After the loop, we immediately check whether we are in position to
provide a final answer:
\lstset{firstnumber=17}
\begin{gospelnumbers}
        if !k = 0 then raise Not_found;
        if 2 * !k > n then raise (Found !cand);
\end{gospelnumbers}
If~\of{k} stores zero, we are sure no candidate has reached majority. We use the
OCaml standard library \of{Not_found} exception to signal such behavior. If the
value stored in~\of{k} is more than half of~\of{n}, the size of the array, we
are sure \of{cand} has reached majority. We use locally-defined exception
\of{Found} to signal this behavior. If none of the above conditions is met, then
we cannot give yet a definitive answer; we need an extra traversal over the
array to check whether \of{cand} has the majority.

The final step of the implementation is a simple loop that counts the actual
number of votes for \of{cand}. If, at some point in the traversal, the
accumulated votes for \of{cand} are greater then half of~\of{n}, we terminate
signaling the majority of this candidate. We can then reuse reference~\of{k} for
the purpose of counting votes:
\lstset{firstnumber=19}
\begin{gospelnumbers}
        k := 0;
        for i = 0 to n - 1 do
         (*@ invariant !k = numof_eq a !cand 0 i && 2 * !k <= n *)
          if E.eq a.(i) !cand then begin
            incr k;
            if 2 * !k > n then raise (Found !cand) end
        done;
        raise Not_found
      with Found c -> c
\end{gospelnumbers}
%
% Mário: explicar convenientemente o invariant, sem entrar em
% considerações se é simples ou não
% The supplied loop invariant is straightforward.
%
This loop invariant states that reference \of{k} stores the actual
number of occurrences of \of{cand} and that, if we keep iterating,
then it must be the case that the \of{k} does not yet represent the
majority of votes.
If we reach past the loop, then \of{cand} does not have the majority,
neither does any other candidate. Once again, we use \of{Not_found} to
signal such an outcome.

The final piece in our verified \ocaml implementation of the
Boyer-Moore algorithm is the actual specification for function
\of{mjrty}. This is as follows:
\lstset{firstnumber=28}
\begin{gospelnumbers}
    (*@ c = mjrty a
          requires 1 <= Array.length a
          ensures  2 * numof_eq a c 0 (Array.length a) >
                   Array.length a
          raises   Not_found -> forall x.
                     2 * numof_eq a x 0 (Array.length a) <=
                     Array.length a *)
  end
\end{gospelnumbers}
As a precondition, we assume that the input array has at least one
element. For the regular postcondition, \emph{i.e.}, the one reached
by catching exception~\texttt{Found}, we prove that the returned candidate
indeed has the absolute majority of votes. Finally, in the exceptional
postcondition (\emph{i.e.}, the one reached by raising exception
\of{Not_found}), we prove that no candidate has enough votes to reach
majority.

\subsubsection{Definition of \of{numof_eq} Function.}
\label{sec:defin-ofnum-funct}

To conclude our proof of \texttt{mjrty} implementation, we must
provide an actual definition for function \of{numof_eq}. We do so by
means of an auxiliary function \of{numof}. This is defined, in
\GOSPEL, as follows:
\begin{gospel}
  (*@ function rec numof (p : integer -> bool) (a b : integer) :
        integer
      = if b <= a then 0 else
        if p (b - 1) then 1 + numof p a (b - 1)
                     else     numof p a (b - 1) *)
  (*@ variant b - a *)
\end{gospel}
The call \of{numof p a b} returns the number of integer values, within
a certain range $[\mathtt{a}; \mathtt{b})$, that satisfy a given
predicate~\of{p}. We attach the variant \of{b - a} to the above
definition, which allows us to prove this is a total function.

To define \of{numof_eq}, we specialize \of{numof} for arrays
and an equality predicate:
\begin{gospel}
  (*@ function numof_eq (a : 'a array) (v : 'a) (l u : integer) :
        integer
      = numof (fun j -> a.(j) = v) l u *)
\end{gospel}
The use of the higher-order function \of{numof} leads to a concise and
elegant definition for \of{numof_eq}. This is in an interesting
application of functional programming concepts to derive sound,
expressive, and yet intuitive specifications even in the presence of
mutable data structures.

The right-to-left definition of \of{numof} is useful when it comes to
proving the preservation of loop invariants where one is scanning an
array from left to right. At the beginning of the $i$-th iteration,
one assumes that \of{numof p 0 i} represents the number of elements,
in the slice $[0; \mathtt{i})$ of some array, that respect
predicate~\texttt{p}. At the end of the iteration, we must
re-establish the invariant for the range $[0; \mathtt{i}+1)$,
\emph{i.e.}, \of{numof p 0 (i + 1)}. If the $i$-th element respects
predicate~\of{p}, then we take the \of{then} branch in the definition
of \of{numof}; otherwise, we take the \of{else} branch. In either
cases, the invariant is re-established simply following the definition
of \of{numof}, since the recursive call \of{numof p 0 i} is exactly
what we assumed at the beginning of the iteration. This applies to the
proof of invariant preservation for the loops in the \of{mjrty}
function, where \of{numof_eq a !cand 0 i} is mapped into a call to
\of{numof (fun j -> a.(j) = !cand) 0 i}.

Finally, we provide auxiliary lemmas about the behavior of the
\of{numof} function that allows us to close the proof of the
\texttt{MJRTY} algorithm. First, we establish the lower and upper
bound for the result of \of{numof}. A call \of{numof a b p}, for any
given integer values~\of{a} and~\of{b} and a predicate~\of{p}, always
returns a non-negative value and cannot exceed \of{b - a}. This is
captured by the following lemma:
\begin{gospel}
  (*@ lemma numof_bounds :
        forall p : (integer -> bool), a b : integer.
        a < b -> 0 <= numof p a b <= b - a *)
\end{gospel}
This lemma is proved interactively by induction on~\of{b}, starting
from~\of{a}.

The next lemma states that a call to \of{numof p a c} can be written
as the sum of calling \of{numof} in the range
$[\mathtt{a}; \mathtt{b}[$ and calling \of{numof} in the range
$[\mathtt{b}; \mathtt{c}[$, provided that
$\mathtt{a} \leq \mathtt{b} \leq \mathtt{c}$. This is expressed as
follows:
\begin{gospel}
  (*@ lemma numof_append:
      forall p: (integer -> bool), a b c: integer.
      a <= b <= c -> numof p a c = numof p a b + numof p b c *)
\end{gospel}
This lemma is proved by induction on~\of{c}, starting from~\of{a}.

The last two lemmas capture what happens in a single computation step
of a call \of{numof p l u}, when \of{l < u}. On one hand, if
value~\of{l} respects predicate~\of{p}, then we add 1 to the result of
the recursive call %
\of{numof p (l + 1) b}:
\begin{gospel}
  (*@ lemma numof_left_add :
        forall p : (integer -> bool), l u : integer.
        l < u -> p l -> numof p l u = 1 + numof p (l + 1) u *)
\end{gospel}
On the other hand, if~\of{l} does not respect~\of{p}, then %
\of{numof p a b} is simply the result of the recursive call:
\begin{gospel}
  (*@ lemma numof_left_no_add:
        forall p : (integer -> bool), l u : integer.
        l < u -> not p l -> numof p l u = numof p (l + 1) u *)
\end{gospel}
One can also think of these lemmas as establishing the equivalence
between either counting the number of elements that satisfy a given
predicate from left-to-right, or from right-to-left. Both lemmas are
proved by instantiating lemma \texttt{numof\_append}, where~\of{a} is
instantiated with~\of{l},~\of{b} with~\of{l + 1}, and~\of{c}
with~\of{u}.

The \cameleer-\why pipeline generates a total of 25 VCs for function
\of{mjrty}. These are discharged using a combination of the
\textsf{Alt-Ergo}, \textsf{Z3}, and \textsf{CVC5} solvers. The proof
of the auxiliary lemmas is also carried in the \why IDE, using
dedicated transformations for induction over integer numbers and
instantiating other lemmas. For reference, the complete \ocaml
implementation, \GOSPEL specification, and auxiliary definitions
% definition of logical functions \of{numof} and \of{numof_eq}, and
% auxiliary lemmas
for the
\texttt{MJRTY} algorithm is given in
Appendix~\ref{sec:boyer-moore-mjrty}.

\subsection{Summary}
\label{sec:summary-1}

In this section, we showed how to use the imperative traits of \ocaml
to write elegant and efficient code. Moreover, with \cameleer, we are
still able to prove the correctness of such implementations. The
gallery of \cameleer verified programs includes several examples of
verified imperative implementations. From those, we highlight the
verification of a \texttt{Union Find} data structure, encoded in an
array of integer values. An important feature of this case study is
the use of the \emph{decentralized invariants}
technique~\cite{filliatre20jlamp} to achieve a fully-automated proof.

\mario{Exemplo grande: union find.}

\mario{Muito boa oportunidade para explicar a técnica de invariantes
  descentralizados.}

\mario{Segmentos de listas? Será que o único exemplo interessante é o da
  \of{Queue}? Que outros exemplos? --- Ver no paper CPP do Arthur.}

\section{Related Work}
\label{sec:related-work}

Deductive software verification is now a mature discipline that is
taught, worldwide, in the vast majority of Computer Science
curricula. However, a large corpus of pedagogical, practical, hands-on
oriented bibliography is still missing. One can cite the recent book
on Dafny~\cite{10.1007/978-3-642-17511-4_20} as valuable contribution
to fill this gap. On the other end of the spectrum of verification
tools, the book by Nipkow \textit{et al.}\cite{nipkow2021functional}
and the Software Foundations volumes 3~\cite{Appel:SF3} and
6~\cite{Chargueraud:SF6} provide comprehensive collections of data
structures and algorithms formally verified in proof assistants. The
first is completely developed in \textsf{Isabelle}, the other two in
\textsf{Coq}.

When it comes to deductive verification of programs written in
functional languages, one can cite frameworks like
\textsf{Iris}~\cite{DBLP:journals/jfp/JungKJBBD18} and Hoare Type
Theory~\cite{DBLP:journals/jfp/NanevskiMB08}. These are built on
\textsf{Coq}, on top of very rich reasoning logics based on Separation
Logic. These can scale up to the verification of complex imperative
and concurrent programs. However, proofs in such frameworks are
conducted manually, requiring a high degree of human interaction and
proof expertise. In the particular case of verification of \ocaml
programs, the \cfml~\cite{chargueraud2011characteristic} tool takes as
input an \ocaml program and translates it into a \textsf{Coq} term
that captures the semantics of the program. The proof is then
conducted using Separation Logic. \cfml proofs are laborious and, in
particular when compared with \cameleer, require extensive human
interaction.

% Closer to our presentation of \ocaml verified programs, one can cite
% the Software Foundations Volume 6 by
% Charguéraud~\cite{Chargueraud:SF6}. This textbook provides a very
% in-depth presentation of Separation Logic inside the \coq proof
% assistant. If features a large set of exercises, ranging from basic
% proofs on references to more advanced higher-order stateful functions
% (\emph{e.g.}, in-place concatenation of linked-lists in
% continuation-passing style). This book is also a comprehensive
% description of the fundamentals underlying the \cfml tool. Contrarily
% to our use of real \ocaml code, the programs verified in this Volume
% are written in a ML-like language, purposely design for this book and
% embedded within \coq.

% \mario{Volume 6 do SF.}

% On the \why proof technology, there are many master courses on formal
% methods and software verification taught using such tool, all around
% the globe. On one hand, one can cite the courses given in French
% Universities by \why
% developers\footnote{\url{https://wikimpri.dptinfo.ens-cachan.fr/doku.php?id=cours:c-2-36-1}}. On
% the other hand, the \why team maintains a list of courses taught
% outside of France\footnote{\url{https://www.why3.org/}}. There is,
% however, no definitive reference book on how to properly use the tool,
% both for writing verified software and how to build verification tools
% that can interface with \why.

\mario{Cursos de verificação em Why3, mais ou menos dispersos.}

\mario{Curso de estruturas de dados funcionais em Isabelle. --- Citar o livro.}

\mario{Talvez agora colocar algo sobre \cfml no related work?}

% Mário: eliminado por sugestão do reviewer 3
% Finally, begin a rather recent development, there is very few available material
% on \cameleer, even less with a pedagogical and teaching purpose. Other than the
% conference publications, the only material on the tool are informal, very
% general presentations and seminars given by the author. This tutorial paper is a
% step forward to remedy this situation, since it gives a transversal overview of
% the tool and its main features.

\section{Conclusions and Future Perspectives}
\label{sec:conclusion}

In this tutorial paper, we presented the deductive verification of
different \ocaml programs, ranging from purely functional
implementations to code combining imperative features, such as mutable
state and local exceptions. We use the \cameleer tool to conduct our
practical experiments. This tool takes as input an actual \ocaml
implementation and translates it into an equivalent \whyml program,
the language of the \why verification framework. \cameleer avoids the
need to re-write entire \ocaml code bases, just for sake of
verification, as it would be the case with a direct use of \why: one
would have first to write a \whyml implementation and specification,
then rely on an extraction mechanism to get an executable equivalent
\ocaml program. On the other hand, \cameleer is conceived with a clear
focus towards proof automation, improving on the experience of
entirely conducting manual proofs in an interactive proof assistant.

% Mário: eliminado por sugestão do reviewer 3
% One distinguishing feature of our tutorial is the incremental
% approach we took: first, only purely functional programs; second, we
% introduce imperative constructions but avoid pointer manipulation;
% finally, we embrace the use Separation Logic to tackle the
% verification of heap-dependent programs. Through the lens of
% classical data structures, this incremental path allowed us to
% revisit many interesting features of the \ocaml language and how
% verification tools incorporate such features. Our effort shows that
% \ocaml is a good vehicle to build correct software, and indeed we
% have the right tools to prove the correctness of \ocaml code.

\mario{Dizer agora que de facto podemos usar o \ocaml como veículo para
  construir software verificado e que temos as ferramentas certas para isso.}

Throughout the paper, we use \GOSPEL to attach formal specification to
\ocaml programs. Our experience suggests that this language is a good
compromise when it comes to conciseness and readability of
specifications, without sacrificing rigor. This is a major argument to
bring even more \ocaml programmers to adopt formal methods techniques
in their daily routines.
Finally, \GOSPEL can be used not only for deductive verification but
also for dynamically analyze \ocaml code. As future work, it would be
interesting to collaboratively use static and dynamic analysis
techniques to tackle the verification of different parts of a big
piece of \ocaml software, resorting to \GOSPEL as the aggregation
entity.
%
% %
% One might say, \textit{different tools, one specification language to
%   rule them all}.

\mario{Novidade: neste tutorial, abordagem pela lente das estruturas de dados
  com abordagem incremental. Utilização conjunta de diferentes ferramentas, mas
  para uma mesma linguagem (\ocaml) e usando uma linguagem de especificação
  comum (\GOSPEL).}

\mario{Dizer aqui que pode ser uma abordagem a seguir num curso de verificação
  dedutiva: functional --> imperative --> separation logic.}

\mario{Diferentes ferramentas --- um mesmo propósito --- one language to rule
  them all.}

\mario{More specification patterns --- iteration.}

% Mário: eliminado por sugestão do reviewer 3
% Finally, we believe this tutorial is an important step forward to produce a
% large body of pedagogical, hands-on material on the deductive verification of
% \ocaml-written code. This would be an important addition to the \ocaml
% community, providing the adequate teaching support on verification for regular
% programmers. In fact, we believe our incremental approach, and the fact we can
% do it all in a single programming language, form a good basis for a whole course
% on deductive verification of functional programs.

\paragraph{Acknowledgments.} I sincerely thank the anonymous
reviewers from the Formal Methods 2024 Tutorial Track. Their comments
and suggestions have greatly improved the presentation of this paper.

\subsubsection{Data Availability.}
\label{sec:data-availability}

The artifact supporting the experiments of this paper is publicly available at
Zenodo, \url{https://doi.org/10.5281/zenodo.12588707}.

\bibliography{local}
\bibliographystyle{splncs04}

\newpage

\appendix

\section{Same Fringe Implementation}
\label{sec:same-fringe-impl}

\begin{gospel}
module type EQUAL = sig
  type t

  val eq : t -> t -> bool
  (*@ b = eq x y
        ensures b <-> x = y *)
end

module Make (E : EQUAL) = struct
  type tree = Empty | Node of tree * E.t * tree

  (*@ function elements (t : tree) : E.t list =
        match t with
        | Empty -> []
        | Node l x r -> (elements l) @ (x :: elements r) *)

  type zipper = (E.t * tree) list

  (*@ function enum_elements (e : zipper) : E.t list =
        match e with
        | [] -> []
        | (x, r) :: e -> x :: (elements r @ enum_elements e) *)

  let rec mk_zipper (t : tree) (e : zipper) =
    match t with
    | Empty -> e
    | Node (l, x, r) -> mk_zipper l ((x, r) :: e)
  (*@ r = mk_zipper t e
        variant t
        ensures enum_elements r = elements t @ enum_elements e *)

  let rec eq_zipper(e1 : zipper) (e2 : zipper) =
    match (e1, e2) with
    | [], [] -> true
    | (x1, r1) :: e1, (x2, r2) :: e2 ->
       E.eq x1 x2 && eq_zipper (mk_zipper r1 e1) (mk_zipper r2 e2)
    | _ -> false
  (*@ b = eq_num e1 e2
        variant List.length (enum_elements e1)
        ensures b <-> enum_elements e1 = enum_elements e2 *)

  let same_fringe (t1 : tree) (t2 : tree) =
    eq_zipper (mk_zipper t1 []) (mk_zipper t2 [])
  (*@ b = same_fringe t1 t2
        ensures b <-> elements t1 = elements t2 *)
end
\end{gospel}

\section{Boyer and Moore MJRTY Algorithm}
\label{sec:boyer-moore-mjrty}

\begin{gospel}
(*@ function rec numof (p: integer -> bool) (a b: integer) :
      integer
    = if b <= a then 0 else
      if p (b - 1) then 1 + numof p a (b - 1)
                   else     numof p a (b - 1) *)
(*@ variant b - a *)

(*@ lemma numof_bounds :
      forall p : (integer -> bool), a b : integer.
      a < b -> 0 <= numof p a b <= b - a *)

(*@ lemma numof_append:
      forall p: (integer -> bool), a b c: integer.
      a <= b <= c -> numof p a c = numof p a b + numof p b c *)

(*@ lemma numof_left_add :
      forall p : (integer -> bool), l u : integer.
      l < u -> p l -> numof p l u = 1 + numof p (l + 1) u *)

(*@ lemma numof_left_no_add:
      forall p : (integer -> bool), l u : integer.
      l < u -> not p l -> numof p l u = numof p (l + 1) u *)

(*@ function numof_eq (a: 'a array) (v: 'a) (l u: integer) :
      integer
    = numof (fun j -> a.(j) = v) l u *)

module type EQUAL = sig
  type t

  val eq : t -> t -> bool
  (*@ b = eq x y
        ensures b <-> x = y *)
end

module Mjrty (E : EQUAL) = struct
  type candidate = E.t

  let mjrty a =
    let exception Found of candidate in
    let n = Array.length a in
    let cand = ref a.(0) in
    let k = ref 0 in
    try
      for i = 0 to n - 1 do
        (*@ invariant 0 <= !k <= numof_eq a !cand 0 i
            invariant 2 * (numof_eq a !cand 0 i - !k) <= i - !k
            invariant forall c. c <> !cand ->
                        2 * numof_eq a c 0 i <= i - !k *)
        if !k = 0 then begin
          cand := a.(i);
          k := 1 end
        else if E.eq !cand a.(i) then incr k
        else decr k
      done;
      if !k = 0 then raise Not_found;
      if 2 * !k > n then raise (Found !cand);
      k := 0;
      for i = 0 to n - 1 do
        (*@ invariant !k = numof_eq a !cand 0 i /\ 2 * !k <= n *)
        if E.eq a.(i) !cand then begin
          incr k;
          if 2 * !k > n then raise (Found !cand) end
      done;
      raise Not_found
    with Found c -> c
  (*@ c = mjrty a
        requires 1 <= Array.length a
        ensures  2 * numof_eq a c 0 (Array.length a) > Array.length a
        raises   Not_found ->
                   forall x. 2 * numof_eq a x 0 (Array.length a)
                          <= Array.length a *)
end
\end{gospel}

\end{document}

%% file: mymacros.tex
\newcommand{\whyml}{\textsf{WhyML}\xspace}

\newcommand{\ocaml}{\textsf{OCaml}\xspace}

\newcommand{\cfml}{\textsf{CFML}\xspace}

\newcommand{\why}{\textsf{Why3}\xspace}

\definecolor{thegray}{rgb}{0.9,0.9,0.9}
\definecolor{colorspec}{rgb}{0,0,0.797}
\definecolor{thered}{rgb}{0.797,0,0}
\definecolor{darkgreen}{rgb}{0.797,0,0}
\definecolor{theblue}{rgb}{0,0,0.797}
\definecolor{darkgray}{rgb}{0.8477,0.8477,0.8477}
\definecolor{ocaml-bg}{rgb}{0.9,0.9,0.9}
\definecolor{thegraygray}{rgb}{0.5,0.5,0.5}